\documentclass[%
 aip,
 amsmath,amssymb,
 reprint,%
]{revtex4-1}
\usepackage{xcolor}
\usepackage{graphicx}% Include figure files
\usepackage{dcolumn}% Align table columns on decimal point
\usepackage{bm}% bold math
\usepackage[utf8]{inputenc}
\usepackage[T1]{fontenc}
\usepackage{mathptmx}
\usepackage{etoolbox}

\makeatletter
\def\@email#1#2{%
 \endgroup
 \patchcmd{\titleblock@produce}
  {\frontmatter@RRAPformat}
  {\frontmatter@RRAPformat{\produce@RRAP{*#1\href{mailto:#2}{#2}}}\frontmatter@RRAPformat}
  {}{}
}%
\makeatother
\begin{document}

\preprint{AIP/123-QED}

\title[Mapping Chaos: Bifurcation Patterns and Shrimp Structures in the Ikeda Map]{Mapping Chaos: Bifurcation Patterns and Shrimp Structures in the Ikeda Map}
% Force line breaks with \\
\author{Diego F. M. Oliveira}
\email{diegofregolente@gmail.com}
\affiliation{ 
School of Electrical Engineering and Computer Science, College of Engineering \& Mines - University of North Dakota, Grand Forks, North Dakota, USA%\\This line break forced with \textbackslash\textbackslash
}%
\date{\today}% It is always \today, today,

\begin{abstract}
This study examines the dynamical properties of the Ikeda map, with a focus on bifurcations and chaotic behavior. We investigate how variations in dissipation parameters influence the system, uncovering shrimp-shaped structures that represent intricate transitions between regular and chaotic dynamics. Key findings include the analysis of period-doubling bifurcations and the onset of chaos. We utilize Lyapunov exponents to distinguish between stable and chaotic regions. These insights contribute to a deeper understanding of nonlinear and chaotic dynamics in optical systems.

\end{abstract}

\maketitle

\begin{quotation}
{The study of dynamical systems is crucial for understanding complex behaviors in various natural and engineered processes. This paper investigates the properties of the Ikeda map, a well-known model in chaos theory that describes the behavior of light in a nonlinear optical cavity. Despite its simplicity, the Ikeda map exhibits a rich variety of dynamical behaviors, including fixed points, periodic orbits, and chaotic attractors. This research focuses on the impact of dissipation parameters on the map's dynamics, demonstrating the existence of self-similar shrimp-shaped structures within the parameter space. These structures delineate regions of stability and chaos, characterized by transitions from regular to chaotic dynamics via a period-doubling bifurcation cascade. The Lyapunov exponent is used as the primary tool to classify regions in the parameter space as either regular or chaotic, revealing the intricate interplay between order and chaos. Through numerical analysis, we also estimate Feigenbaum's constant, further validating the observed bifurcation patterns. Our findings contribute to a deeper understanding of the complex parameter space of the Ikeda map, highlighting its significance in the broader context of nonlinear dynamical systems.}
\end{quotation}

\section{\label{sec:level1} Introduction}

Dynamical systems are crucial for understanding the complex behaviors that arise in various natural and engineered processes. Among the most extensively studied dynamical systems are the Lorenz system \cite{lorenz1963deterministic}, the Hénon map \cite{gallas1993structure}, the logistic map \cite{may1976simple,strogatz2018nonlinear}, and the Duffing oscillator \cite{guckenheimer2013nonlinear,moon1979magnetoelastic}. These nonlinear systems, along with many others, display a rich spectrum of behaviors ranging from regular and predictable to chaotic and unpredictable dynamics. The phase space of such systems can generally be divided into three regions: regular \cite{kamphorst1999bounded,oliveira2010dynamical,oliveira2012scaling}, chaotic \cite{pustyl1995construction,sinai1970dynamical,bunimovich1979ergodic,robnik1983classical,oliveira2011fermi,oliveira2012flight}, and mixed regions \cite{leonel2005hybrid,oliveira2009scaling,oliveira2011boundary,oliveira2015symmetry,page2020iris}.

Regular regions are characterized by periodic or quasi-periodic trajectories that repeat over time, leading to stable and predictable behavior. Chaotic regions, in contrast, exhibit sensitive dependence on initial conditions, where even minute differences in starting points result in vastly divergent trajectories, leading to an unpredictable and complex phase space. Mixed regions are particularly intriguing as they encompass both regular and chaotic dynamics, with stable islands of regularity embedded within chaotic seas, making the overall behavior of the system highly intricate.

In this paper, we explore the properties of a dynamics of the Ikeda map \cite{ikeda1979multiple,ikeda1980optical,watanabe1994constants}.  Introduced by Kensuke Ikeda in the late 1970s, the Ikeda map models the behavior of light in a nonlinear optical cavity. It has since become a quintessential example in the study of chaos and complex systems. 
The Ikeda map is defined by a set of iterative equations that describe the evolution of a point in the complex plane. Despite its relatively simple form, the Ikeda map exhibits a rich variety of dynamical behaviors, including fixed points, periodic orbits, and chaotic attractors. The map is particularly notable for its sensitivity to parameter changes, which can lead to sudden transitions between regular and chaotic dynamics as we will show below. Additionally, as we will demonstrate, increasing the dissipation parameter leads to a period-doubling bifurcation cascade \cite{linsay1981period}, where Feigenbaum's constant, \(\delta\) \cite{feigenbaum1978quantitative,feigenbaum1979universal,hanias2009period,chen2012new}, which quantifies the rate of these bifurcations, can be calculated numerically. We then examine the two-dimensional parameter space, specifically focusing on the dissipation parameters associated with the real and imaginary components of the map, and the model reveals the presence of self-similar structures known as ``shrimps".

Initial investigations by Gaspard et al. \cite{gaspard1984bifurcation} in 1984, Rössler et al. \cite{rossler1989modulated} in 1989, and Komuro et al. \cite{komuro1991global} in 1991 laid the groundwork for understanding self-similar periodic structures in two-dimensional mappings. Gaspard and colleagues focused on Chua's system, revealing complex bifurcation patterns, while Rössler's study on the logistic map and Komuro's exploration of the Double Scroll circuit further illustrated the ubiquity of these structures in nonlinear dynamical systems.

\begin{figure*}[t]
\begin{center}
\centerline{(a)\includegraphics[width=8.0cm,height=6.cm]{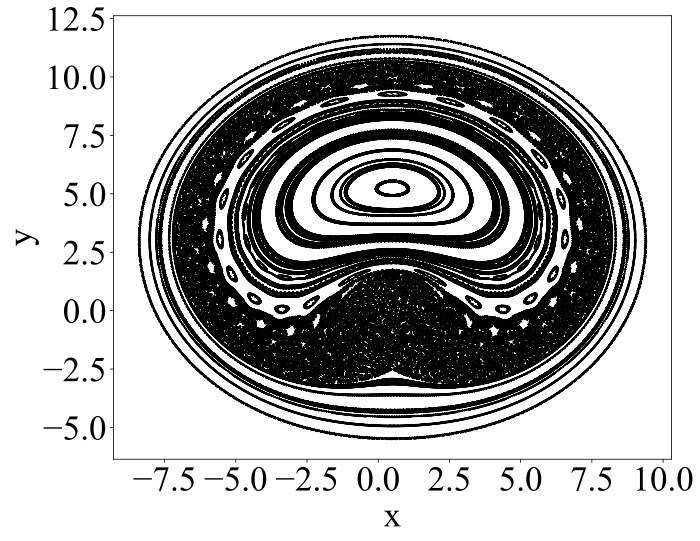}
			(b)\includegraphics[width=8.0cm,height=6.cm]{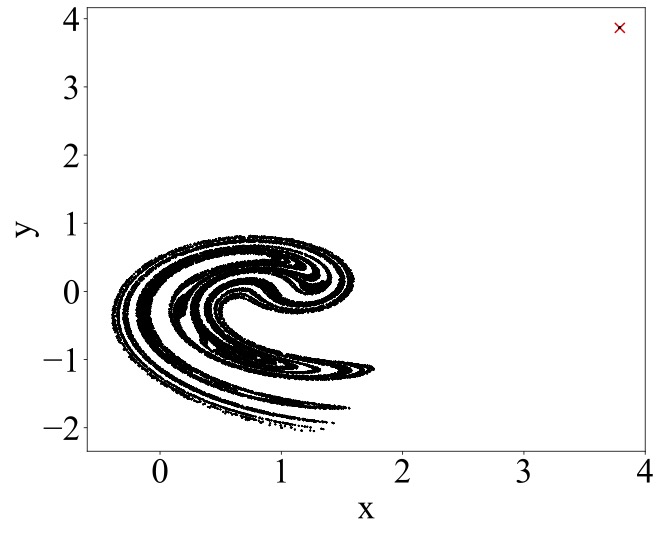}}
\centerline{(c)\includegraphics[width=8.0cm,height=6.cm]{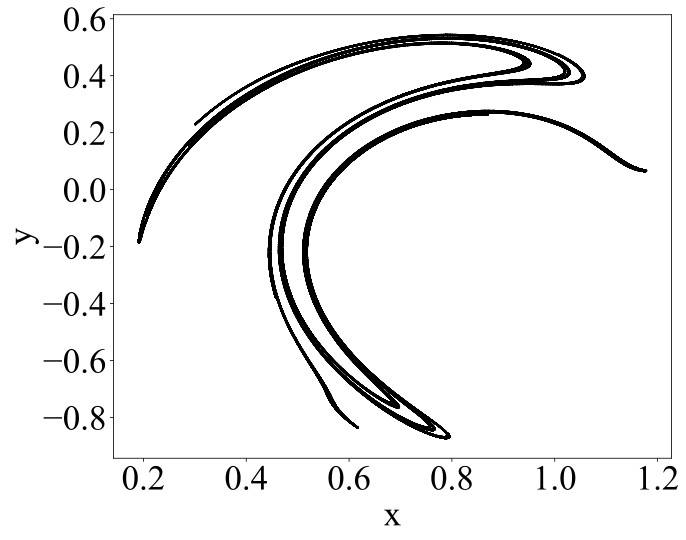}
			(d)\includegraphics[width=8.0cm,height=6.cm]{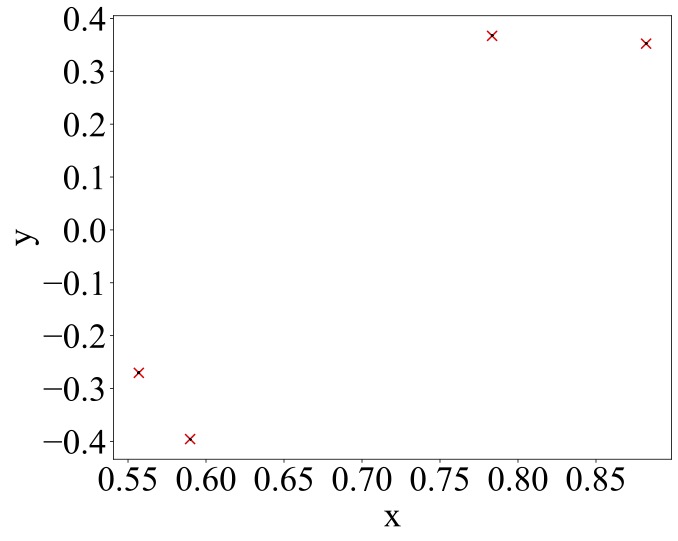}}
\end{center}   
\caption{Phase space for the Ikeda map. The figures were constructed using the following control parameters: (a) \(u_x=1\) and \(u_y=1\), which correspond to conservative dynamics; (b) \(u_x=0.95\) and \(u_y=0.85\), where a chaotic attractor and an attracting fixed point are observed; (c) \(u_x=0.7\) and \(u_y=0.8\), where only a chaotic attractor is present; and (d) \(u_x=u_y=0.6\), where a period-4 attracting fixed point is observed.}
\label{fig1}
\end{figure*}

Gallas's groundbreaking work in 1993 \cite{gallas1993structure} represented a pivotal moment in the study of dynamical systems. Through a detailed exploration of the parameter space of the Hénon map, Gallas identified the presence of complex shrimp-shaped domains—regions characterized by periodic behavior amidst chaotic dynamics. This discovery not only highlighted the significance of these structures but also spurred a wave of subsequent research focused on uncovering similar patterns in other models. As noted by \cite{vitolo2011global}, "Shrimps are formed by a regular set of adjacent windows centered around the main pair of intersecting superstable parabolic arcs. A shrimp is a doubly infinite mosaic of stability domains, comprising an innermost main domain and all adjacent stability domains arising from two period-doubling cascades and their corresponding chaos regions. It is important to distinguish these shrimp from their innermost main domain of periodicity."  Since Gallas's seminal work, shrimp-shaped domains have been recognized in a wide array of theoretical models, including but not limited to those examined in studies by Gallas himself \cite{gallas1995structure,gallas1994dissecting}, as well as in investigations by Hunt et al. \cite{hunt1999bifurcation}, Bonatto et al. \cite{bonatto2005self}, and Oliveira et al. \cite{oliveira2013some,oliveira2012dynamical,oliveira2014statistical}. 
Furthermore, in addition to theoretical explorations, E. N. Lorenz \cite{lorenz2008compound}, one of the foremost pioneers of chaos theory, extended the investigation of shrimp-shaped domains into new territories, focusing on their appearance in complex systems. Lorenz's work highlighted the richness and complexity of these structures, solidifying their importance in the study of chaotic dynamics.
These theoretical studies have contributed to a deeper understanding of the conditions under which these domains emerge and their implications for the dynamics of the systems in which they appear. More recently, the existence of shrimp-shaped domains has been experimentally verified in physical systems, most notably in electronic circuit \cite{maranhao2008experimental,stoop2010real,stoop2012shrimps,viana2010high}. These experimental observations have bridged the gap between theory and practice, demonstrating that these intricate structures are not merely mathematical curiosities but are also observable in real-world systems.

To classify regions in the parameter space as either regular or chaotic, we primarily use the Lyapunov exponent. Although other quantities, such as particle velocity, can also be employed—like in the well-known Fermi-Ulam model \cite{diego2011parameter}—they may not effectively distinguish between regular and chaotic behavior. In this study, we focus on the Lyapunov exponent. Our approach involves starting with a fixed initial condition, allowing for an extended transient period, and then computing the Lyapunov exponent. For each parameter combination of the dissipation paramenters, we assign a color based on the computed Lyapunov exponent. We then increment the parameters, using the final values of the dynamical variables before the increment as the new initial condition, ensuring that we remain within the basin of the same attractor. This methodology reveals well-organized, self-similar shrimp-shaped structures embedded within a broader region of chaotic attractors.

\section{The Model and the Map}

The Ikeda map, originally introduced to model the dynamics of a laser system, is described by the following 1D equation involving a complex variable:
\begin{equation}
z_{n+1} = A + B z_n  e^{i \left( \theta - \frac{\phi}{|z_n|^2 + 1} \right)},
\label{Eq001}
\end{equation}
where the parameters \(\theta\) and \(\phi\) typically represent phase angles that control the behavior of the system,  \( z_n = x_n + i y_n \) is a complex number with real part \( x_n \) and imaginary part \( y_n \), and \( |z_n|^2 = x_n^2 + y_n^2 \) is the squared magnitude of \( z_n \). In order to obtain the two-dimensional map the describes the dynamics of the system, first, one compute the exponential term in Eq. \ref{Eq001}.Thus, by using Euler's formula, we have:

\begin{equation}
e^{i \left( \theta - \frac{\phi}{|z_n|^2 + 1} \right)} = \cos\left(\theta - \frac{\phi}{|z_n|^2 + 1}\right) + i \sin\left(\theta - \frac{\phi}{|z_n|^2 + 1}\right).
\end{equation}

Substitute \( z_n = x_n + i y_n \) and the exponential term into the map:

\begin{equation}
z_{n+1} = A + B (x_n + i y_n) \left[\cos\left(\theta - \frac{\phi}{x_n^2 + y_n^2 + 1}\right) + i \sin\left(\theta - \frac{\phi}{x_n^2 + y_n^2 + 1}\right)\right] e^C.
\end{equation}

Expanding and separating the real and imaginary parts, we obtain:
\begin{widetext}
\begin{eqnarray}
z_{n+1} = A &+& B e^C \left\{\left[ x_n \cos\left(\theta - \frac{\phi}{x_n^2 + y_n^2 + 1}\right) - y_n \sin\left(\theta - \frac{\phi}{x_n^2 + y_n^2 + 1}\right)\right] + 
  i \left[x_n \sin\left(\theta - \frac{\phi}{x_n^2 + y_n^2 + 1}\right) + y_n \cos\left(\theta - \frac{\phi}{x_n^2 + y_n^2 + 1}\right) \right]\right\}.
\end{eqnarray}
\end{widetext}
Assuming that \( A = 1 \) and \( B e^C = u_i \), with $i=x$ for the real component or $i=y$  for the imaginary componedt of the mapping and defining \( t_n = \theta - \frac{\phi}{x_n^2 + y_n^2 + 1} \). Then the map can be simplified as:

\begin{equation}
z_{n+1} = 1 + u_i \left[ (x_n \cos t_n - y_n \sin t_n) + i (x_n \sin t_n + y_n \cos t_n) \right].
\end{equation}

\begin{figure}[t]
\centerline{\includegraphics[width=9.0cm,height=8cm]{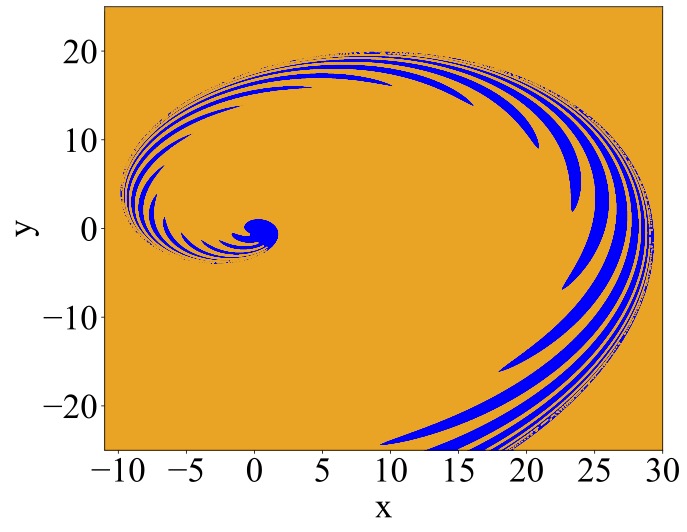}}
\caption{(Color online) Basing of attraction for the chaotic attractor (blue) and the attracting fixed point (orange) shown in Figure \ref{fig1} (c).}
\label{fig2w}
\end{figure}

Finally, by separating the real and imaginary parts of \( z_{n+1} \), we obtain the discrete two-dimensional non-linear mapping that describes the system's dynamics:

\begin{eqnarray}
S:\left\{\begin{array}{ll}
x_{n+1} &= \text{Re}(z_{n+1}) = 1 + u_x (x_n \cos t_n - y_n \sin t_n), \\
y_{n+1} &= \text{Im}(z_{n+1}) = u_y (x_n \sin t_n + y_n \cos t_n),
\end{array} ~,
\right.
\label{map2d}
\end{eqnarray}
where \( t_n = \theta - \frac{\phi}{x_n^2 + y_n^2 + 1} \) and \( u_x \in [0,1] \) and \( u_y \in [0,1] \) are the dissipation factors for the real and imaginary components of the map. From now on, we will fix $\theta=0.4$ and $\phi=0.6$. These factors modulate the transformation of the real and imaginary parts of the complex variable through each iteration. This transformation effectively maps the original complex map into a system of coupled real-valued equations, which can be analyzed to study the dynamical behavior of the system. It is important to
mention that if $u_x=u_y=1$
all the results for the Hamiltonian area-preserving  map are recovered. From the map $S$ [see Eqs. (\ref{map2d})], one can easily
obtain the Jacobian matrix, $J$, which is
defined as
\begin{equation}
J = \begin{pmatrix}
\frac{\partial x_{n+1}}{\partial x_n} & \frac{\partial x_{n+1}}{\partial y_n} \\
\frac{\partial y_{n+1}}{\partial x_n} & \frac{\partial y_{n+1}}{\partial y_n}
\end{pmatrix}
\label{eq7rsm}
\end{equation}
with coefficients given by the following expressions

\begin{eqnarray}
\frac{\partial x_{n+1}}{\partial x_n} &=& u_x \left[ \cos t_n - (x_n \sin t_n + y_n \cos t_n) \frac{\partial t_n}{\partial x_n} \right] \\
\frac{\partial x_{n+1}}{\partial y_n} &=& u_x \left[ -\sin t_n - (x_n \sin t_n + y_n \cos t_n) \frac{\partial t_n}{\partial y_n} \right] \\
\frac{\partial y_{n+1}}{\partial x_n} &=& u_y \left[ \sin t_n + (x_n \cos t_n - y_n \sin t_n)\frac{\partial t_n}{\partial x_n} \right] \\
\frac{\partial y_{n+1}}{\partial y_n} &=& u_y \left[ \cos t_n + (x_n \cos t_n - y_n \sin t_n)\frac{\partial t_n}{\partial y_n} \right]
\end{eqnarray}
where $\frac{\partial t_n}{\partial x_n}$ and $\frac{\partial t_n}{\partial y_n}$ are given by

\[
\frac{\partial t_n}{\partial x_n} = \frac{2\phi x_n}{(1 + x_n^2 + y_n^2)^2}, \quad \frac{\partial t_n}{\partial y_n} = \frac{2\phi y_n}{(1 + x_n^2 + y_n^2)^2}
\]
After some calculation one can show that the map is area preserving only when $u_x=u_y=1$ since the determinant of the Jacobian matrix is $\det(J)=u_xu_y$. As an ilustration, Figure \ref{fig1} illustrates the structure of the phase space for the map \ref{map2d} with varying values of \( u_x \) and \( u_y \). Considering $u_x=u_y=1$ [ Fig. \ref{fig1}(a)], we observe a chaotic sea interspersed with a set of Kolmogorov-Arnold-Moser (KAM) islands. In this scenario, the phase space exhibits regions of chaotic behavior, where trajectories appear to move unpredictably and ergodically. Amidst this chaotic sea, KAM islands emerge as stable, quasi-periodic regions where the motion is regular and confined. These islands represent the remnants of the invariant tori that survive the perturbation introduced by the system's non-linearity. The coexistence of chaotic regions and KAM islands highlights the complex and rich structure of the phase space, illustrating the intricate interplay between order and chaos in the dynamical system. As dissipation is introduced, the phase space structure undergoes significant changes. The system exhibits a complex interplay between chaotic and periodic dynamics, as demonstrated in Fig. \ref{fig1}(b), where both chaotic and periodic orbits coexist. Specifically, Fig. \ref{fig1}(b) illustrates an attractive fixed point (indicated by a \textcolor{red}{$\times$}) alongside a chaotic attractor, each with its own basin of attraction, as further detailed in Fig. \ref{fig2w}. As the system parameters vary, the dynamics undergo significant changes. For instance, in Fig. \ref{fig1}(c), a chaotic attractor emerges, signaling a transition from periodic to chaotic behavior. However, with further decreases in the dissipation parameters, the chaotic attractor is eventually replaced by a set of attracting fixed points, as shown in Fig. \ref{fig1}(d). This progression underscores the system's intricate and diverse behavior as it navigates different regions of parameter space. Figure \ref{fig2w} presents the basins of attraction for both the attracting fixed point (orange) and the chaotic attractor (blue) depicted in Fig. \ref{fig1}(b). To generate this figure, the ranges \( u_x \in [-11, 30] \) and \( u_y \in [-25, 25] \) were partitioned into grids of 2000 intervals each, resulting in a total of \( 4 \times 10^6 \) distinct initial conditions. Each initial condition was iterated up to \( n = 5 \times 10^5 \). While other attractors may exist, their basins of attraction, if present, are either too small to detect or lie outside the scope of the initial conditions considered in this study.

\begin{figure}[htb]
\centerline{(a)\includegraphics[width=8.0cm,height=5cm]{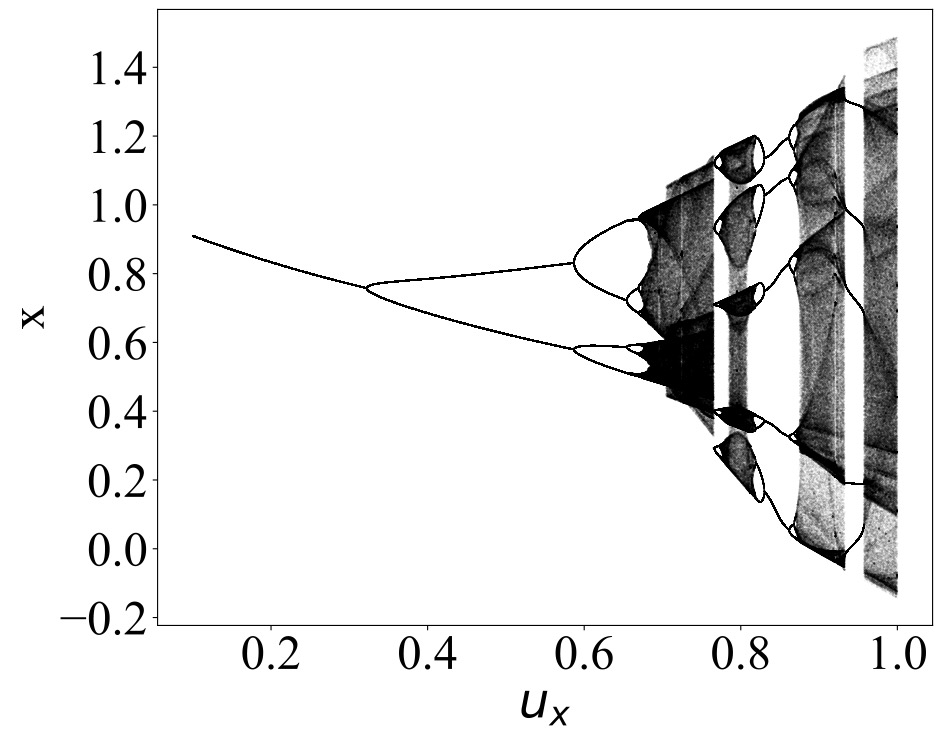}}
\centerline{(b)\includegraphics[width=8.0cm,height=5cm]{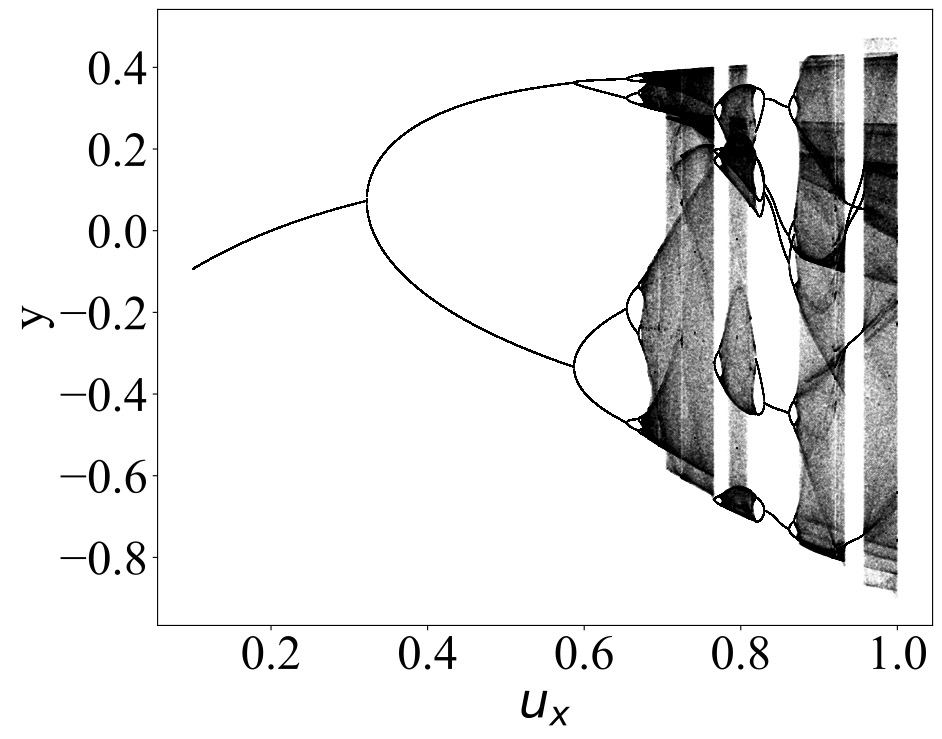}}
\centerline{(c)\includegraphics[width=8.0cm,height=5cm]{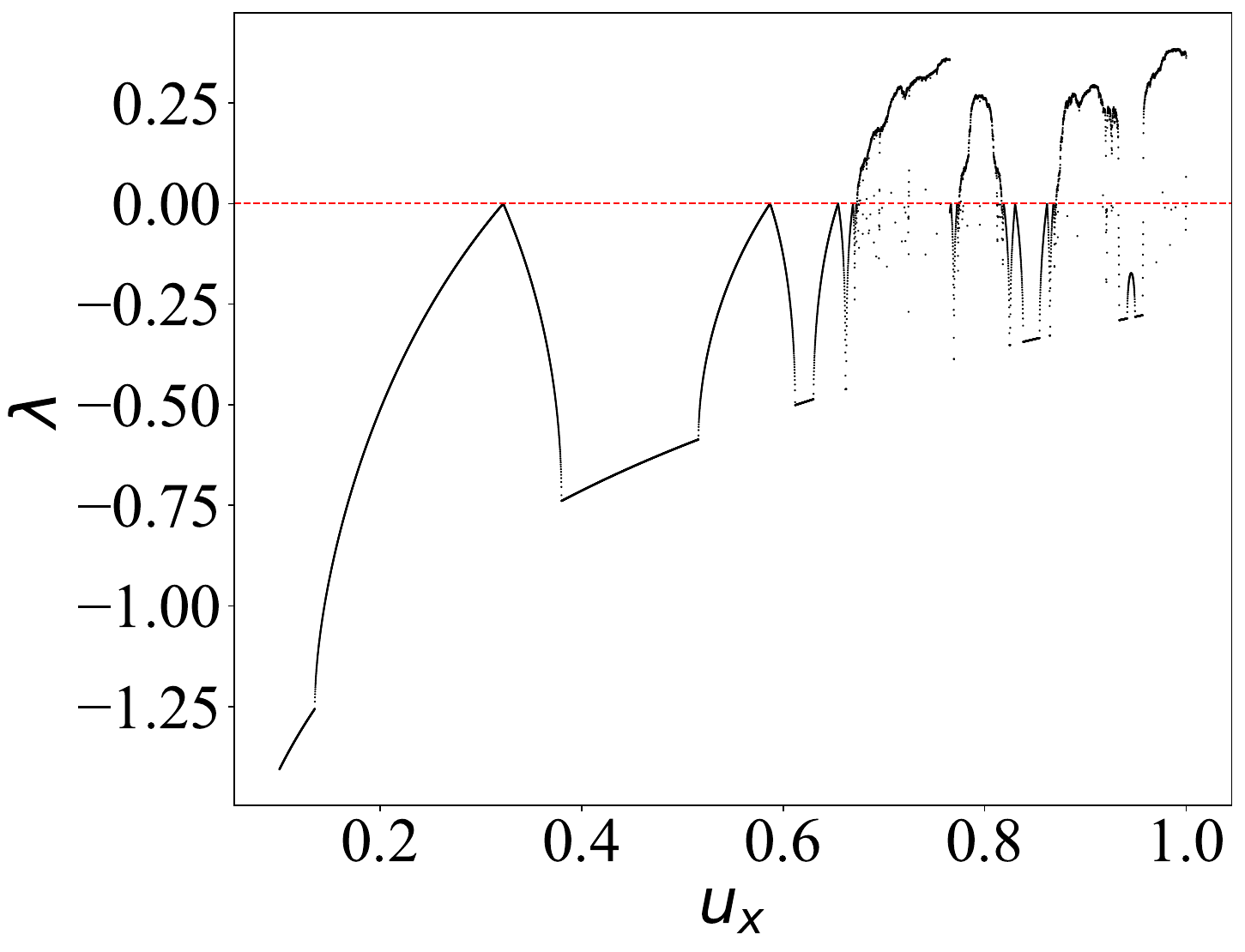}}
\caption{(Color online) Bifurcation cascade for (a) \(x\) and (b) \(y\) plotted against the parameter \(u_x\) with \(u_y\) fixed at 0.6. In (c), the Lyapunov exponent corresponding to (a) and (b) is shown. The red dotted line indicates where the Lyapunov exponent is equal to zero.
}
\label{fig3}
\end{figure}

Furthermore, as we have shown, by slightly changing the control parameter, the behavior of the initial conditions can shift from chaotic to periodic.  In Fig. \ref{fig3} (a-b), we see that for lower values of \( u_x \), the system tends to exhibit periodic behavior, characterized by regular oscillations of \( x \) and \( y \). As \( u_x \) increases, these periodic regions become interspersed with chaotic intervals, where the trajectories of \( x \) and \( y \) become irregular and sensitive to initial conditions. The transition between these behaviors is marked by bifurcations, indicating changes in the system's stability and the emergence of new dynamical regimes. We can observe the transition from regular to chaotic behavior by examining the bifurcation diagram. In this analysis, we consider the case where \( u_y = 0.6 \). To explore typical behaviors, specifically the bifurcation diagrams as the control parameter \( u_x \) varies, we use the initial conditions \( x_0 = 0.2 \) and \( y_0 = 0.3 \). Figure \ref{fig3}(a) shows the behavior of \( x \) plotted against the control parameter \( u_x \), where a sequence of period-doubling bifurcations is evident. A similar sequence is observed for the asymptotic variable \( y \), as shown in Figure \ref{fig3}(b). It is important to note that the bifurcations of the same period in both (a) and (b) occur for the same values of the control parameter \( u_x \).  Feigenbaum \cite{feigenbaum1978quantitative,feigenbaum1979universal} was the first to observe a ``universal'' feature in the behavior of bifurcations in dynamical systems. Specifically, he noticed that as a system transitions to chaos through a series of period-doubling bifurcations, the ratios of the differences in control parameter values at which these bifurcations occur converge geometrically at a constant rate, denoted as \(\delta\). This discovery indicates that there is a universal behavior in a wide range of systems approaching chaos.

\newcounter{rownum}
\setcounter{rownum}{0}
% Table
\begin{table}[t]
\begin{center}
\begin{tabular}{|p{.80cm}||p{1.0cm}|p{3.2cm}|p{3.cm}|}
\hline
\textbf{~~~~${\bf n}$} & \textbf{$~{\bf period}$} & \textbf{$~~~~~~~~~~~~~~~~~~{\bf  u_x}$}  & \textbf{$~~~~~~~~~~~~~~~~{\bf \delta}$} \\ \hline \hline
%Increment then display the counter value
~~~~~{\bf 1} & ~~~~~2 & ~0.32219134277600003  &  ~~~~~~~~~------\\ \hline
~~~~~{\bf 2} & ~~~~~4 & ~0.58704813585000004  & ~3.217851540432144 \\ \hline
~~~~~{\bf 3} & ~~~~~8 & ~0.66935671651200002  & ~26.06556657140821  \\ \hline
~~~~~{\bf 4} & ~~~~16 & ~0.67251446796299996  & ~4.656191581014719  \\ \hline
~~~~~{\bf 5} & ~~~~32 & ~0.67319265127999994  & ~4.668065705181482  \\ \hline
~~~~~{\bf 6} & ~~~~64 & ~0.67333793272200004  & ~4.668654531417602  \\ \hline
~~~~~{\bf 7} & ~~~128 & ~0.67336905120400004  & ~4.668788977835374  \\ \hline
~~~~~{\bf 8} & ~~~256 & ~0.67337571641893001  & ~4.669542136198597 \\ \hline
~~~~~{\bf 9} & ~~~512 & ~0.67337714379975999  & ~{\bf 4.6692}48396257327  \\ \hline
~~~~{\bf 10} & ~1024 & ~0.67337744949795997  & ~  ~~~~~~~~~------\\ \hline
\end{tabular}
\caption{Shows the value of $n$, the period of the bifurcation, the
values of the parameter $u_x$ where the bifircation happen and the
convergence of the Feigenbaum's $\delta$ considering bifurcations up to
the tenth order.}
\end{center}
\label{tablelabel}
\end{table}%

The procedure to determine the Feigenbaum constant \(\delta\) is methodical: (a) Identify the bifurcation points: Let \(u_x(1)\) be the control parameter value at which a period-1 orbit (a single stable cycle) bifurcates into a period-2 orbit (a stable cycle that repeats every two periods);
    (b) {Continue the process:} Let \(u_x(2)\) be the value where the period-2 orbit bifurcates into a period-4 orbit, and \(u_x(3)\) where the period-4 orbit bifurcates into a period-8 orbit, and so on;
    (c) {Generalize the parameter values:} In general, the parameter \(u_x(n)\) corresponds to the control parameter value at which a period-\(2^n\) orbit is born.
The Feigenbaum constant \(\delta\) is then defined as the limit of the ratio of successive differences between these control parameter values as \(n\) approaches infinity. Mathematically, it is expressed as:

\begin{equation}
\delta = \lim_{n \rightarrow \infty} \frac{u_x(n) - u_x(n-1)}{u_x(n+1) - u_x(n)}.
\label{delta}
\end{equation}
This constant \(\delta\) captures the rate at which the bifurcations occur and is found to be the same for a wide variety of dynamical systems, highlighting the universality of this behavior. The theoretical value of the Feigenbaum constant \(\delta\) is approximately \(4.669201609\ldots\). This value has been confirmed through both numerical calculations and experimental observations, and it plays a crucial role in the understanding of the transition to chaos in nonlinear dynamical systems. How can we determine the parameter where bifurcation occurs? One effective tool is the Lyapunov exponent. As discussed by Eckmann and Ruelle \cite{eckmann1985ergodic}, the Lyapunov exponents are defined as:
\begin{equation}
\lambda_j=\lim_{n\rightarrow\infty}{1\over{n}}\ln|\Lambda_j|~~,~~j=1,
2~~,
\label{eq4}
\end{equation}
where $\Lambda_j$ are the eigenvalues of
$M=\prod_{i=1}^nJ_i(x_i,y_i)$ and $J_i$ is the Jacobian matrix
evaluated over the orbit $(x_i,y_i)$. However, a direct
implementation of a computational algorithm to evaluate Eq. (\ref{eq4})
has a severe limitation to obtain $M$. For the limit of short $n$,
the components of $M$ can assume different orders of magnitude for
chaotic orbits and periodic attractors, making  the implementation of
the algorithm impracticable. To avoid such a problem, $J$ can be
written as $J=\Theta T$ where $\Theta$ is an orthogonal matrix and $T$
is a right up triangular matrix. $M$ is rewritten as $M=J_nJ_{n-1}\ldots
J_2\Theta_1\Theta_1^{-1}J_1$, where $T_1=\Theta_1^{-1}J_1$. A product of
$J_2\Theta_1$ defines a new $J_2^{\prime}$. In a next step, one can
show that $M=J_nJ_{n-1}\ldots J_3\Theta_2\Theta_2^{-1}J_2^{\prime}T_1$.
The same procedure can be used to obtain $T_2=\Theta_2^{-1}J_2^{\prime}$
and so on. Using this procedure, the problem is reduced to evaluate the
diagonal elements of $T_i:T_{11}^i,T_{22}^i$. Finally, the Lyapunov
exponents are given by
\begin{equation}
\lambda_j=\lim_{n\rightarrow\infty}{1\over{n}}\sum_{i=1}^n
\ln|T_{jj}^i|~~,~~j=1,2~~.
\label{eq005}
\end{equation}
If at least one of the $\lambda_j$  is positive, the orbit is said to be chaotic. Figure (\ref{fig3})(c) shows the behavior of the Lyapunov exponents corresponding to Fig. \ref{fig3}(a-b). It is evident that when bifurcations occur, the exponent $\lambda_j$ vanishes. 
Based on the numerical data obtained through the calculation of Lyapunov exponents, Feigenbaum's $\delta$ for the Ikeda map is found to be \(\delta = 4.669248396257327\ldots\). Here, we have considered bifurcations up to the tenth order (see Table I), and our result is in good agreement with Feigenbaum’s \(\delta\) up to \(10^{-4}\).

\begin{figure*}[t]
\begin{center}
\centerline{(a)\includegraphics[width=9.0cm,height=8cm]{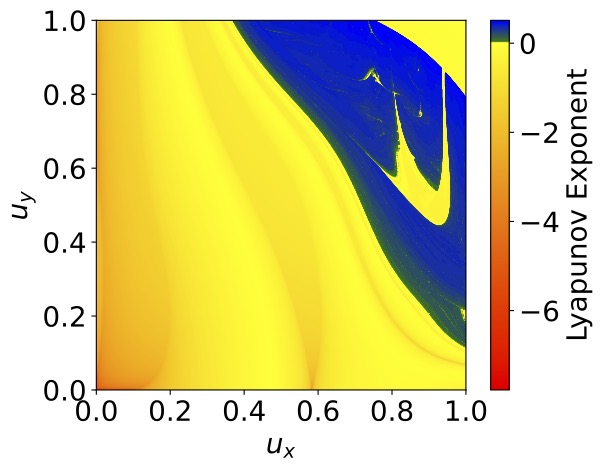}
			(b)\includegraphics[width=9.0cm,height=8cm]{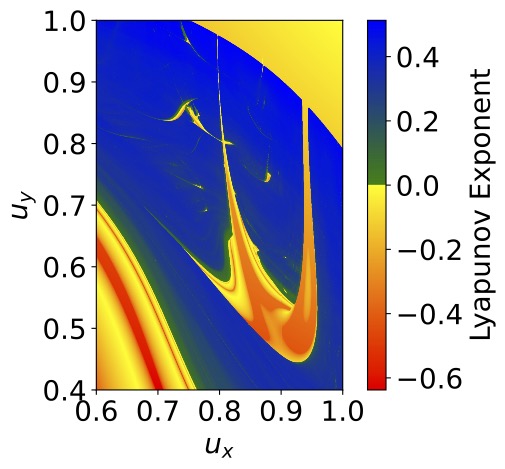}}
% \centerline{(a)\includegraphics[width=9.0cm,height=8cm]{shrimp1.pdf}
% 			(b)\includegraphics[width=9.0cm,height=8cm]{shrimp3.pdf}}  
\end{center}   
\caption{(Color online) (a) Phase diagram of \(u_x\) versus \(u_y\), illustrating the distinct shrimp-shaped structures that characterize different dynamic behaviors. The diagram uses a color scale to represent the Lyapunov exponent for each pair of \((u_x, u_y)\) values. Regions with regular, periodic behavior are highlighted in colors ranging from red to yellow, indicating negative Lyapunov exponent. In contrast, regions exhibiting chaotic dynamics are shown in colors ranging from green to blue, corresponding to positive Lyapunov exponent. Here, one can clearly differentiate between areas of stability and chaos within the parameter space depending on the values of \((u_x, u_y)\). (b) Magnification of the main structure in (a).}
\label{fig2x}
\end{figure*}

Furthermore, the detailed examination of these bifurcation diagrams reveals the underlying mechanisms driving the system towards chaos. Overall, these bifurcation diagrams serve as a crucial tool for visualizing and understanding the complex dynamics of the system as it responds to changes in the control parameter \( u_x \). The clear sequence of period-doubling bifurcations leading to chaos underscores the rich and intricate nature of the system's behavior under the given conditions. 

To obtain a more comprehensive understanding of the model's dynamics, we investigate the parameter space where both components of the dissipation parameters, \( u_x \) and \( u_y \), vary. By systematically varying \( u_x \) and \( u_y \), we can map out the regions of the parameter space that correspond to different dynamical behaviors. This includes identifying zones of periodic motion, chaotic behavior, and the transitions between them. Such a parametric study provides insights into the robustness of the observed phenomena and helps to uncover the underlying structure of the phase space.

The detailed exploration of the parameter space also reveals how the interplay between the dissipation parameters \( u_x \) and \( u_y \) influences the overall dynamics. For example, increasing \( u_x \) might enhance the stability of periodic orbits, while varying \( u_y \) could affect the onset of chaos. By understanding these relationships, we can better predict the system's response to changes in the control parameters and potentially develop strategies to control or exploit the dynamics for specific applications.

To thoroughly investigate the parameter space of the system described in equation \ref{map2d}, we systematically varied the dissipation parameters \( u_x \) and \( u_y \). For each combination of these parameters, after discarding a significant transient period, we calculated the Lyapunov exponent, which is a crucial measure of the system's sensitivity to initial conditions. The value of the Lyapunov exponent determines whether the system exhibits chaotic or regular behavior.

Once the Lyapunov exponent was computed for each pair \( (u_x, u_y) \), we assigned a corresponding color to visually represent the stability or chaos of the system within the parameter space. Specifically, Figure \ref{fig2x} provides a detailed view of the parameter space for the Ikeda map, where a shrimp-shaped structure becomes evident. The color scale used in this figure is carefully designed to distinguish between different types of behavior: regions exhibiting regular, stable dynamics are colored in shades of red to yellow, while regions displaying chaotic behavior are colored in shades of green to blue. This shrimp-shaped structure is a well-known feature in the study of dynamical systems and indicates the presence of complex bifurcation patterns.

Our findings in this parameter space are in agreement with previous studies, particularly with the results presented in Refs. \cite{gallas1993structure,baptista1997parameter,rossler1989modulated,mackay1986transition,celestino2014stable}. To construct the figure, we divided the range of both \( u_x \) and \( u_y \) into 2000 equal intervals, resulting in a grid of \( 4 \times 10^6 \) distinct parameter combinations. For each combination, we began with initial conditions \( x_0 = 0.2 \) and \( y_0 = 0.3 \), and then iteratively followed the attractor as the parameters \( u_x \) and \( u_y \) were incrementally adjusted. After each increment, the final state of the previous combination served as the new initial condition, ensuring continuity in the exploration of the attractor's evolution.

However, it is important to note that this approach can inadvertently omit information about other potential attractors. This omission occurs because the chosen initial conditions might lie within the basin of attraction of a specific attractor, thereby excluding the exploration of other possible attractors that might exist for different initial conditions.

Figure \ref{fig2x}(b) offers a magnified view of the main structure in Figure \ref{fig2x}(a), revealing the intricate bifurcation patterns characteristic of each shrimp-shaped structure. Each shrimp features a central body, generally representing stable dynamics, followed by an infinite sequence of bifurcations that adhere to the pattern \( k \times 2^n \), where \( k \) denotes the period of the central body. These bifurcations give rise to increasingly complex and chaotic dynamics, showcasing the rich and multifaceted behavior of the Ikeda map as the parameters are varied. The detailed bifurcation sequences provide valuable insights into the system’s transition from regular to chaotic dynamics, highlighting the underlying complexity and sensitivity of the parameter space.

\section{Conclusion}

This study explores the rich dynamical behavior of the Ikeda map, a nonlinear system originally designed to model light in a nonlinear optical cavity. Despite its seemingly simple form, the Ikeda map exhibits a complex array of dynamical phenomena, including periodic orbits, chaotic attractors, and intricate bifurcation structures.

Through our analysis, we have demonstrated that the Ikeda map features a diverse range of dynamical regimes depending on the values of the dissipation parameters \(u_x\) and \(u_y\). The phase space analysis reveals a striking transition from chaotic to periodic behavior as these parameters vary. Specifically, the map exhibits a period-doubling bifurcation cascade, characteristic of systems approaching chaos. We have successfully quantified the Feigenbaum constant \(\delta\) associated with these bifurcations, finding it to be consistent with the theoretically established value of approximately \(4.669201609...\). This confirms the universal nature of the bifurcation cascade observed in the Ikeda map and aligns with the behaviors seen in other dynamical systems.

Furthermore, the identification of shrimp-shaped structures within the parameter space provides a compelling visualization of the map's dynamical complexity. These structures represent intricate domains of stability interspersed with chaotic regions, underscoring the interplay between order and chaos in the map's dynamics. Our use of Lyapunov exponents as a tool for distinguishing chaotic from regular regions has proven effective, offering a clear method for analyzing the map's behavior.

Overall, this work highlights the Ikeda map's utility as a model for studying chaotic systems and bifurcation phenomena. The observed transitions between periodic and chaotic behaviors, along with the rich parameter space structures, illustrate the map's capability to capture the essence of dynamical complexity. Future research could extend these findings by exploring other parameter ranges or by examining the influence of additional nonlinearities, further enhancing our understanding of chaos and stability in nonlinear dynamical systems.

\section{ACKNOWLEDGMENTS}

I would like to dedicate this work to the memory of Prof. Dr. Jason Alfredo Carlson Gallas, with whom I had the privilege of meeting during my time as a Postdoctoral Researcher  at the Institute of Multiscale Simulation, Friedrich-Alexander-Universität Erlangen-N\"urnberg, under the supervision of Prof. Dr. Thorsten P\"oschel in 2012.

\section{References}
%\nocite{*}
%\bibliography{aipsamp}% Produces the bibliography via BibTeX.

%

\end{document}